\documentclass{aa}  
\usepackage{graphicx}
\usepackage{txfonts}
\usepackage{color}
\usepackage{longtable}
\begin{document}

   \title{Orbit determination of Transneptunian objects and Centaurs \\ for the prediction of stellar occultations}

   \author{J.~Desmars
          \inst{1,2}
   \and	
	J.I.B.~Camargo\inst{1,3}
   \and	
        F.~Braga-Ribas\inst{1,3,4}
   \and
        R.~Vieira-Martins\inst{1,3}\fnmsep\thanks{Affiliated researcher at Observatoire de Paris/IMCCE, 77 Avenue
Denfert Rochereau 75014 Paris, France}
   \and
        M.~Assafin\inst{5}
   \and
        F.~Vachier\inst{2}
   \and
        F.~Colas\inst{2}
   \and
        J.~L.~Ortiz\inst{6}
   \and
        R.~Duffard\inst{6}
   \and
        N.~Morales\inst{6}
   \and
        B.~Sicardy\inst{7}
   \and
        A.R.~Gomes-J\'{u}nior\inst{5}
   \and
        G.~Benedetti-Rossi\inst{1}
          }

   \institute{Observat\'{o}rio Nacional/MCTI, Rua Gal. Jos\'e Cristino 77, CEP 20921-400, Rio de Janeiro, Brazil
              \and
              Institut de M\'ecanique C\'eleste et de Calcul des \'Eph\'em\'erides - Observatoire 
              de Paris, UMR 8028 CNRS, 77 avenue Denfert-Rochereau, 75014 Paris, France
              \and
              Laborat\'{o}rio Interinstitucional de e-Astronomia - LIneA, Rua Gal. Jos\'e Cristino 77, CEP 20921-400, Rio de Janeiro, Brazil
              \and
              Federal University of Technology - Paran\'a (UTFPR / DAFIS), Rua Sete de Setembro, 3165, CEP 80230-901, Curitiba, PR, Brazil
              \and
              Observat\'{o}rio do Valongo/UFRJ, Ladeira do Pedro Ant\^{o}nio 43, CEP 20080-090, Rio de Janeiro,
              Brazil
              \and
              Instituto de Astrof\'{\i}sica de Andaluc\'{\i}a, CSIC, Apartado 3004, 18080 Granada, Spain
              \and
              LESIA, Observatoire de Paris, CNRS UMR 8109, Universit\'{e} Pierre et Marie Curie, 
              Universit\'{e} Paris-Diderot, 5 place Jules Janssen, F-92195 Meudon Cedex, France
             }

\offprints{J.Desmars, desmars@imcce.fr}
   \date{}

 
  \abstract
   {The prediction of stellar occultations by Transneptunian objects (TNOs) and Centaurs is a difficult challenge that requires accuracy both in the occulted star position as for the object ephemeris. Until now, the most used method of prediction involving tens of TNOs/Centaurs was to consider a constant offset for the right ascension and for the declination with respect to a reference ephemeris, usually, the latest public version. This offset is determined as the difference between the most recent observations of the TNO and the reference ephemeris. This method can be successfully applied when the offset remains constant with time, i.e. when the orbit is stable enough. In this case, the prediction holds even for occultations to occur several days after the last observations.}
   {This paper presents an alternative method of prediction based on a new accurate orbit determination procedure, which uses all the available positions of the TNO from the Minor Planet Center database plus sets of new astrometric positions from unpublished observations.}
   {The orbit determination is performed through a numerical integration procedure called NIMA, in which we develop a specific weighting 
scheme that takes into account the individual precision of observation, the number of observations performed during one night in a same observatory and the presence of systematic errors in the positions.}
   {The NIMA method was applied for 51 selected TNOs and Centaurs. For this purpose, we have performed about 2900 new observations 
in several observatories (European South Observatory, Observat\'orio Pico dos Dias, Pic du Midi, etc) during the 2007-2014 
period. Using NIMA, we succeed in predicting the stellar occultations of 10 TNOs and 3 Centaurs between July 2013 and February 2015. By comparing the NIMA and JPL ephemerides, we highlighted the variation of the offset between them with time, showing that in general the constant offset hypothesis is not valid, even for short time scales of a few weeks. Giving examples, we show that the constant offset method could not accurately predict 6 out of the 13 observed positive occultations successfully predicted by NIMA. The results indicate that NIMA is capable of efficiently refine the orbits of these bodies.
Finally, we show that the astrometric positions given by positive occultations can help to further refine the orbit of the TNO and consequently the future predictions. We also provide the unpublished observations of the 51 selected TNOs and their ephemeris in a usable format by the SPICE library.
}
   {}

   \maketitle
%

\section{Introduction}
When a Transneptunian object (TNO) or a Centaur occults a
star, their sizes and shapes can be determined with kilometric accuracy \citep{2010Natur.465..897E,2011Natur.478..493S, 2012Natur.491..566O,2013ApJ...773...26B} from the resulting light curves of ground-based observers located inside the shadow path or nearly outside it. 
Such an accuracy in dimensions can only be rivaled by space missions. Also, ring systems \citep{2014Natur.508...72B} and atmospheres as tenuous as few nanobars \citep{2009Icar..199..458W,2011Natur.478..493S,2013ApJ...773...26B} can be detected when present. These parameters are important for the study of TNOs\footnote{TNO will indicate both Transneptunian objects and Centaurs hereafter. Centaur objects have their origin often related to the Transneptunian objects.} and, as a consequence, to retrieve the history and evolution of the outer solar system. However, previous to the observation of such an occultation, its prediction (where and when, on Earth, the event can be detected) is essential.

Prediction of stellar occultations requires both accurate position of the occulted star and of the TNO ephemeris. The budget uncertainty of star position and ephemeris must be smaller than the apparent angular size of the body radius, in order for the occultation to be observable at the forseen location on Earth. The position of the star is initially taken from an astrometric catalogue, taking into account its proper motion, if available. As the proper motion and the astrometry may be inaccurate, the star is observed before the occultation in order to refine its position. 
Depending on the number and quality of the observations, the accuracy of the final position is about is about 10 mas to 20 mas which is similar to the apparent angular size of the objects. 

The ephemeris is obtained after an orbit determination process. It consists in the determination of orbital elements that minimise the O-C \textit{i.e.} the difference between the observed and the computed positions. The ephemeris remains precise during the observational period but starts to diverge after the last observations. 

Previous studies about predictions of occultations tried to overcome this ephemeris divergence problem by using more recent observations.  
\cite{Assafin2012,Camargo2014} used a constant offset to the ephemeris to refine the predictions. The offset is determined thanks to a set of observations performed from a few months up to a few weeks before the predicted occultation. In practice, the offset is computed as the difference between the observed position and a reference ephemeris at the date of the offset observations. This 
method assumes that the offset remains constant or varies by less than the body radius until the occultation date, which is only the case when the occultation 
occurs few days after the offset observations or when the orbit is relatively well determined. 

\cite{Fraser2013} highlighted that the offset is not constant and could lead to unreliable predictions. They proposed to 
refine the orbit using the offset observations and adopted a maximum-likelihood approach to correct the orbital elements. 
They used only their own observations and not directly the past observations on Minor Planet Center database. As a consequence, 
their new orbit strongly depends on their new observations. As so, their ephemeris is likely to diverge in a time span shorter than desired.

Finally, the MIT group (Bosh et al., in prep.) also provides predictions of stellar occultations based on a drift and a periodic term for the offset\footnote{\url{http://occult.mit.edu/research/occultationPredictions.php}}.

In this paper, we present a new numerical procedure called NIMA (Sect.~\ref{S:NIMA}) that computes the orbits of TNOs and Centaurs. NIMA consists of a complete process of orbit determination that profits from all available observations of the TNO/Centaur (past observations from MPC, offset observations and unpublished observations, see Sect.~\ref{S:obs}) and a specific weighting scheme for observations (Sect.~\ref{Ss:weight}). We present in Sect.~\ref{S:results} the results of the use of NIMA for a set of 51 TNOs and Centaurs selected for their physical, observable and dynamical caracteristics. We conclude (Sect.~\ref{S:conclusion}) that NIMA is capable of furnishing significantly improved ephemeris for TNOs and Centaurs, allowing for more accurate stellar occultation predictions within a more extended time span, more than any of the other methods used so far.


\section{NIMA ephemeris}\label{S:NIMA}
The NIMA\footnote{Numerical Integration of the Motion of an Asteroid} ephemeris was originally developed for orbit determination of 
Near-Earth Asteroids in the context of the detection of Yarkovsky effect \citep{Desmars2015}. It can be used to determine 
the orbit of any asteroid and, given the similar dynamical conditions, of any TNO or Centaur (see Sect.~\ref{Ss:numint}). 

The NIMA code allows orbit determination and propagation thanks to a numerical integration of the equations of motion. 
Compared with the version applied to NEAs or with other codes for orbit determination such as OrbFit\footnote{The package 
OrbFit is available on \url{http://adams.dm.unipi.it/~orbmaint/orbfit/}}, the main difference is on the weighting scheme 
which is adapted specifically to the prediction of occultations in order to get short term accurate orbit.

The following sections describe the dynamical model, the fitting process and the weighting scheme. 

\subsection{Numerical integration}\label{Ss:numint}
The dynamical model of a TNO's motion includes the gravitational pertubations of the Sun and the eight planets. All planets 
are considered as point masses and the Earth Moon system is considered as a point mass located at the Earth-Moon Barycentre. 
No other perturbations are required since TNOs are distant objects. For example, by adding the three biggest asteroids (Ceres, 
Vesta and Pallas) or by adding Pluto in the dynamical model, we noticed only insignificant changes in orbit determination. The 
masses of the eight planets and their positions are given by JPL ephemeris DE431 \citep{DE431}.    

The equations of motion are numerically integrated throught a Gauss Radau integrator \citep{Everhart1985}. The equations of 
variation as described in \cite{Lainey2004a} are also integrated in order to determine the partial derivatives of the position 
and the velocity components related to the components of state-vector $(c_j)$ which encompasses the position and the velocity vectors at 
a specific epoch. 

\subsection{Fitting process}
The fitting process consists in the determination of six parameters $C=(c_j)$ (the state vector) that minimises the residuals 
$\Delta Y$ (the difference between observed and computed positions). This determination makes use of a Levenberg-Marquardt 
algorithm by iterative corrections of each component of the state-vector. For each iteration, the corrections to apply are 
determined thanks to the partial derivatives (represented by matrix $A$) and the least square method (LSM) \citep[for more 
information, see for example][]{Desmars2009b}. 

\begin{equation}
 (\widehat{\Delta C})= (A^T V_{obs}^{-1} A)^{-1}A^T V_{obs}^{-1} \Delta Y
\end{equation}

In the LSM, a weighting matrix $V_{obs}$ is required and we specifically discuss of the weighting scheme in the next section. 

The normal matrix $N$ and covariance matrix $\Lambda_0$ are defined as: $N=A^T V_{obs}^{-1} A$ and $\Lambda_0=N^{-1}$. \\

The standard deviation of each parameter $(c_j)$ is given by the root square of the diagonal elements of $\Lambda_0$. Moreover, 
the covariance matrix can be linearly propagated at any date $t$ thanks to the equation:

\begin{equation}
 \Lambda(t)= A(t) \Lambda_0  A(t)^{T} 
\end{equation}

where $A(t)$ is the matrix of partial derivatives at date $t$. Thus this linear relation can provide the estimated precision 
of the position in the celestial sphere (right ascension and declination) at any date.

\subsection{Weighting scheme}\label{Ss:weight}
Observed positions have various accuracies and can be correlated. In this context, we have to consider the covariance matrix of the observed positions
 $V_{obs}$ that is supposed to be known in the LSM. In practice we neglect correlations, so that the covariance matrix of the observed positions 
$V_{obs}$ is considered as a diagonal matrix where the diagonal components are $\epsilon_i^2=1/\sigma_i^2$ and $\sigma_i^2$ 
is the estimated variance of the observed position $i$ \citep{Desmars2009b}. \\ 

In orbit determination, the main difficulty is to give an appropriate weight to each position. 
Positions do not contain only a random error but also many systematic errors with several different sources such as the 
telescope used for observation, the stellar catalogue used for the reduction, etc. 

\cite{Carpino2003} and \cite{Chesley2010} discussed weighting schemes and showed the orbit determination improvement by 
weighting positions according to the observatory and the stellar catalogue used for the reduction. 

However, a problem can appear using this weighting scheme when several dozens of observations were performed during a same 
night in the same observatory. If individual observed positions have a weight $\epsilon_i$, the mean weight for the set of positions 
will be $\epsilon_i/\sqrt{N}$ where $N$ is the number of positions per night. The mean weight can become small whereas positions 
can be biased. In that case, orbit determination will be degraded by such a set of positions.
This problem mainly concerns our offset observations (see Sect.~\ref{Ss:obsoffset}) with an average of 13.1 observations per night whereas in MPC observations (see Sect.~\ref{Ss:obsMPC}), there are 2.8 observations per night for the studied objects. 

To overcome with this problem, we have adopted a specific weighting scheme by taking into account the estimated precision of each 
position depending of each observatory and stellar catalogue used but also a possible bias due to the observatory. 

The estimated variance $\omega_i$ of each position $i$ is given by: 
\begin{equation}
	 \omega_i^2=N_i b_i^2+\sigma_i^2
\end{equation}
where $N_i$ is the number of observations performed during the same night and in the same observatory than position $i$, 
$b_i$ corresponds to the possible bias depending on the stellar catalogue and observatory and $\sigma_i$ is the estimated 
precision of individual position $i$ provided by \cite{Chesley2010} or by Table \ref{T:sigma} for some specific cases.

This weighting scheme avoids assigning an artificial strong weight for a night with several dozens of observations. In such a 
case, the mean variance tends to $b_i^2$ and not $0$. Another interpretation is that we consider that the mean variance for 
a single night cannot be smaller than $b^2_i$. 

Estimated bias $b_i$ and precision $\sigma_i$ will depend on the type of positions we deal with. For example, positions 
from MPC will be considered as average positions since the process of reduction for each position is not completely known. 
On the contrary, our positions performed to determine the offset, will be considered as precise positions since we know 
exactly how they were reduced and the quality of the stellar catalogue used for reduction.

We have performed many tests to determine appropriate bias and precision for the different types of positions (see Sect.~\ref{S:obs}). Finally, 
we empirically adopt the values given in Table~\ref{T:bias} for the bias $b_i$ and the values given in Table~\ref{T:sigma} for 
the individual precision of each position $\sigma_i$. For positions from MPC, we adopt $b_i=300$ mas and $\sigma_i$ 
depends on stellar catalogue and observatory and is given by \cite{Chesley2010}.  

\begin{table*}[h!]
\begin{center}
\caption{Estimated bias for offset observations in both right ascension and declination for different stellar catalogues and observatories}\label{T:bias}
\begin{tabular}{llclr}
\hline
\hline
Source 			& Catalogue 	& IAU code  & Observatory 			& estimated bias\\
  			&       	&  	    &  		 			& $b_i$ (in mas)\\
\hline	
\hline	
MPC			& ---		& ---	&	all				&	300  \\
\hline
Offset			&	WFI	& 809	&	European South Observatory	&	75  \\
			&		& 874	& 	Observat\'orio do Pico dos Dias	&	150  \\
			&		& 586	& 	Pic du Midi			&	150  \\
			&		& 493	& 	Calar Alto			&	150  \\
			&		& J86	&	Sierra Nevada			&	150  \\
			&		& I95	&	La Hita				&	150  \\
			&		& Z20	&	Mercator La Palma		&	150  \\
			&		& J13	&	Liverpool La Palma		&	150  \\
			&		& --- 	&	Other				&	300  \\
\hline
			& UCAC/other	& 809	&	European South Observatory	&	150  \\
			&		& 874	& 	Observat\'orio do Pico dos Dias	&	225  \\
			&		& 586	& 	Pic du Midi			&	225  \\
			&		& 493	& 	Calar Alto			&	225  \\
			&		& J86	&	Sierra Nevada			&	225  \\
			&		& I95	&	La Hita				&	225  \\
			&		& Z20	&	Mercator La Palma		&	225  \\
			&		& J13	&	Liverpool La Palma		&	225  \\
			&		& --- 	&	Other				&	300  \\
\hline
Occultation		& ---		& 244	& 	Geocenter			&	0  \\
\hline
\cite{Fraser2013} 	& 2MASS/SDSS	& 267	& 	CFHT	&	300	\\
		 	& 		& 568	& 	Gemini-Mauna Kea			&	300	\\
\hline
\hline
\end{tabular}
\end{center}
\end{table*}

\begin{table*}[h!]
\begin{center}
\caption{Estimated precision for offset observations in both right ascension and declination for different observatories}\label{T:sigma}
\begin{tabular}{llcr}
\hline
\hline
Source 			&  IAU code  & Observatory 			& estimated precision \\
 			&  	  & 	 			&  $\sigma_i$ (in mas)\\
\hline	
\hline	
MPC			& ---	&	all				&	\cite{Chesley2010} \\
\hline
Offset			& 809	&	European South Observatory	&	bias  \\
			& 874	&	Observat\'orio do Pico dos Dias &	bias  \\
			& 586	&	Pic du Midi			&	bias  \\
			& 493	&	Calar Alto			&	bias  \\
			& J86	&	Sierra Nevada			&	bias  \\
			& I95	&	La Hita 			&	bias  \\
			& Z20	&	Mercator La Palma		&	bias  \\
			& J13	&	Liverpool La Palma		&	bias  \\
			& ---	&	Other				&	bias  \\
\hline
Occultation		& 244	&	Geocenter -accurate position-	&	40  \\
			&  	&	Geocenter -approximate position-&	75  \\
\hline
\cite{Fraser2013} 	& 267	& CFHT	&	bias	  \\
		 	& 568	&	Gemini-Mauna Kea			&	bias	  \\
\hline
\hline
\end{tabular}
\end{center}
\end{table*}

As a comparison, \cite{Fraser2013} used an uncertainty of 40 and 80~mas for their positions whereas the average value for a position from MPC in in AstDys database is about 0.5~arcsec. Consequently, positions from \cite{Fraser2013} have a weight 100 times more 
important than an average position, which seems not appropriate. 

In our study, the maximum precision for a series of several positions in one single night is given by the bias. In this context, 
our best positions that come from ESO and reduced with WFI catalogue (described below), will have a weight about 50 times better than an average 
position from MPC. Positions from ESO and reduced with UCAC4 will have a weight 10 times better than average positions.

\section{Astrometric observations} \label{S:obs}
For TNOs and Centaurs, most of the positions come from Minor Planet Center database. But since about 2007 we also started to observe these objects, in order to check occultation predictions. Obviously, the first computed information to evaluate the orbit status was the average offset between the observed positions and the object ephemeris. For this reason, here and throughout the paper, these kind of observations are called offset observations. Finally, we also used astrometric positions deduced from previous positive occultations.

\subsection{MPC observations}\label{Ss:obsMPC}
Minor Planet Center\footnote{\url{http://www.minorplanetcenter.net/}} (MPC) is in charge of receiving and distributing the positional measurements of minor planets, comets and outer irregular natural satellites. For one specific object, the MPC gives the file of observations, the orbital elements, the ephemeris, and many other data. Observations are provided by many different observers and observatories (from professional to amateur telescopes) and positions are derived from a variety of reference catalogues and position reduction procedures. Consequently, the quality of the observations is heterogeneous and that is why the use of a weighting scheme in orbit determination, taking into account the quality of positions, is important. 

Due to average quality and precision, most of the positions on the MPC database are provided with 5 decimal digits in time that correspond to less that one second of uncertainty which is enough for TNO/Centaurs, with 2 decimal digits in right ascension that corresponds to 150 mas of uncertainty, and with 1 decimal digit in declination that corresponds to 100 mas of uncertainty. Clearly, if this uncertainty is enough for problem of identification and ephemeris of position, it is not for the stellar occultations context that requires a precision of, at least, 50 mas. However, recent observations are sometimes provided with 1 extra decimal digit in time, right ascension and declination.

\subsection{Offset observations} \label{Ss:obsoffset}
A useful step in the process of predicting stellar occultations to a large
number of TNOs, for a span of some few years, is to properly determine a set of 
initial predictions. This set of initial predictions must be complete, that is, 
must contain all possible events of a given TNO involving stars up to a given 
magnitude, and must be accurate enough to allow for a selection of those ones for 
which observational efforts to refine the initial prediction are worth employing.

All initial predictions of stellar occultations to the objects presented in 
Table~\ref{T:residuals}, exception made to (2060) Chiron and (60558) Echeclus, 
are detailed in \cite{Assafin2012} and \citep{Camargo2014} and were based on observations made at La Silla (Chile)
with the ESO 2.2m telescope equipped with the Wide Field Imager (WFI).

The observational runs had two different purposes: to cover the future sky path 
of a given TNO and then to observe the TNO itself. The first set of observations
aimed at determining candidate stars to be occulted by the TNO. As a by product
of the respective observations, catalogues with positions and proper motions
were created to serve as an accurate and - most important - dense reference
catalogue for astrometry. These were called WFI catalogues. The second set of 
observations were used to obtain positions of the TNOs and to determine the 
respective corrections to their ephemerides. The WFI catalogues were used
as reference for the astrometry of these images. (2060) Chiron and (60558) 
Echeclus were included later in our list of objects. Candidate stars to be occulted 
by them were selected from UCAC4 \citep{UCAC4} and USNO-B1 \citep{USNOB} catalogues and consequent
observations to refine the position of the candidate stars and to correct
their ephemerides were carried out at the Pico dos Dias observatory.

For the initial predictions given by both \cite{Assafin2012} and \citep{Camargo2014}, corrections to ephemerides 
were done by considering an average offset as determined from the position differences in the sense "observation minus ephemeris".
Such correction (offset), as mentioned earlier in the text, 
was assumed to be constant until a new one was determined with newer observations.

Although good results were obtained with this method, thanks to the efforts of many
observers, the use of offsets is not ideal (see, for instance, Fig.~\ref{F:2004NT33}).
This method does not allow for predictions with accuracy better than 30 mas, on the position of the TNO, with an advance greater than six months. Orbit re-determination is a more straightforward solution, and the offset observations are still useful to refine the orbit.

\subsection{Astrometry from occultations}\label{Ss:occ}
Once a stellar occultation is detected, besides the physical properties of the object, we can determine its position on the sky relative to the occulted star. From a multi-site observation, the limb fit of the object to the observed occultation light-curves (chords), has as by-product the centre of the body. This position is relative to a given position, usually the occulted star position, so the position of the object's centre at the middle of the event can be directly calculated, with kilometric accuracy.

The absolute position of the object is then dependent (and limited) on the accuracy of the star position, which is usually many times greater than the lib-fit errors. We performed observations of the occulted stars
at Pico dos Dias Observatory to reduce this source of uncertainty.
The observations were made near the epoch of the occultation, usually used to update the predictions, so proper motion errors were avoided. Astrometric reductions were made using the WFI catalogue when available, otherwise the UCAC4 catalogue was used as reference. Obtained accuracies on the star positions are on the order of 10 mas to 20 mas, which is a about the apparent angular size of the TNOs on our list.  

Accurate TNO positions from stellar occultation can be obtained even for single-chord detections. In this case, it is not possible to derive the size of the object, but we can use its estimated size\footnote{Sizes of objects are from \url{http://www.johnstonsarchive.net/astro/tnodiam.html}.} to derive its centre. The observed chord is fitted to a presumed circular object and the centre is calculated with respect to the star position. This will lead to a determination of the centre with an error of few hundreds of kilometres (about the precision of the object size), even considering a north or south solution. The error on the absolute object centre is still dominated by the absolute star position.

This is a very straightforward way to obtain precise TNO positions and was applied to all the detected stellar occultations by TNOs \citep{braga-ribas14b}. This will be specially interesting when GAIA catalogue is available, as the position of the objects will no longer be limited by the accuracy of the star position.

\section{Results}\label{S:results}
\subsection{Comparison between ephemerides}
Ephemerides for TNOs can be found on several databases. To make comparisons with the NIMA ephemeris, we take on the main databases: 
JPL Horizons\footnote{\url{http://ssd.jpl.nasa.gov/horizons.cgi}}, VO-Miriade\footnote{\url{http://vo.imcce.fr/webservices/miriade/}}, 
and AstDys\footnote{\url{http://hamilton.dm.unipi.it/astdys/index.php}} (AstDys makes use of the Orbfit package). Minor Planet Center 
also provides ephemeris but we did not consider it because the coordinate values are truncated to 0.1s in right ascension and 1 arcsec
in declination, which is clearly not enough for predictions of occultations.

\begin{figure}[h!] 
\centering 
\includegraphics[width=\columnwidth]{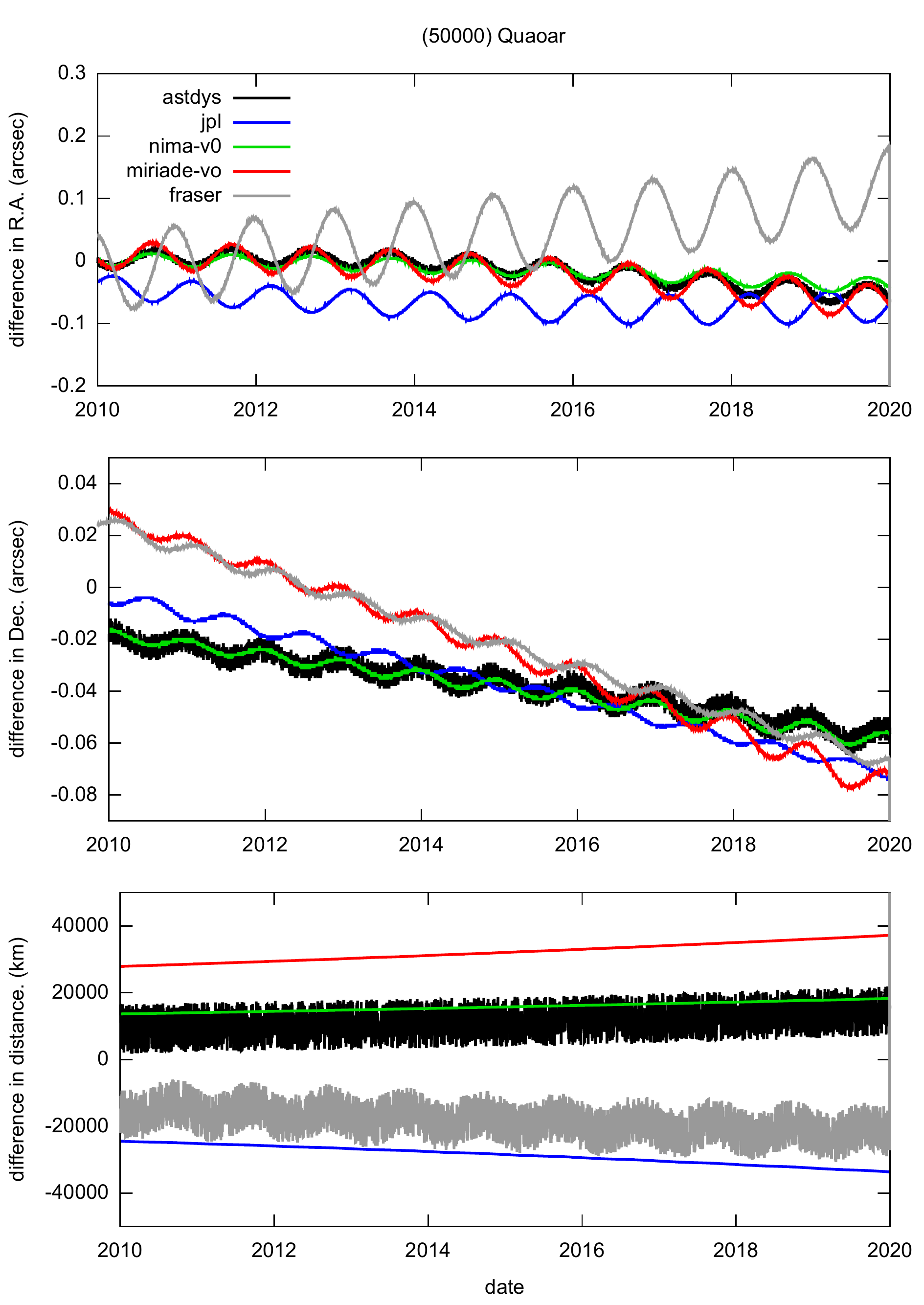}
\caption{Difference in right ascension weighted by $\cos \delta$, declination and geocentric distance between NIMA and each ephemeris for (50000) Quaoar.}\label{F:diffQuaoar}
\end{figure}

\begin{figure}[h!] 
\centering 
\includegraphics[width=\columnwidth]{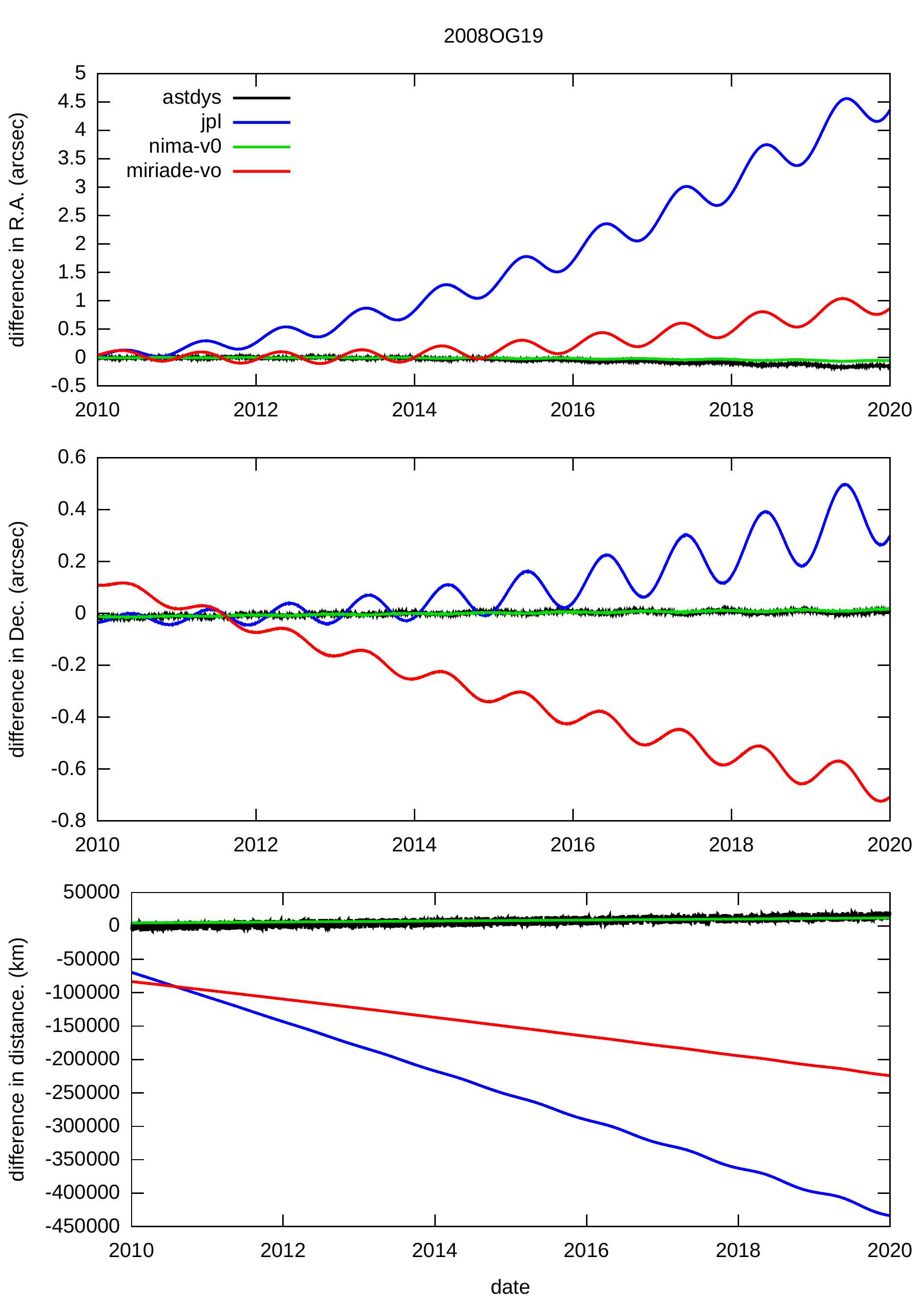}
\caption{Difference in right ascension weighted by $\cos \delta$, declination and geocentric distance between NIMA and each ephemeris for 2008OG19.}\label{F:diff2008OG19}
\end{figure}

Assuming that these databases were produced by using the same set of positions available on the MPC, we can say that they fitted their orbits with the same set of positions as we used with NIMA. 
We compared the ephemeris for two specific objects: (50000) Quaoar which has a long period of observations (1954-2014) and 2008OG19 with a small observational period (2008-2012). 
Figures \ref{F:diffQuaoar} and \ref{F:diff2008OG19} present the difference between the different ephemerides in right 
ascension\footnote{The difference in right ascension is weighted by $\cos \delta$.}, declination, and geocentric distance during 2010-2020 
period. We also use a version of NIMA (nima v0) without changing the weight of positions, i.e. using the weight of \cite{Chesley2010} as in Orbfit. Moreover for Quaoar, we also add the ephemeris from \citep{Fraser2013} just for information because the ephemeris can not be fully compared with the other ones since it made use of additional observations published in the same paper.
 
For objects with a large period of observations such as Quaoar, the differences between the ephemerides are small (less than 0.1~arcsec) 
on the 2010-2020 period. Even with Fraser ephemeris, the difference is small whereas they used additional observations. As they used OrbFit package for orbit determination, the difference without these additional observations, would probably be close to AstDys ephemeris. The small differences between ephemerides for Quaoar indicate the good quality of the orbit, because of the long period of observations available. For objects with a short period of observations such as 2008OG19, the differences between ephemerides are larger (several arcseconds) indicating the low-quality of the orbit. 

The difference between ephemerides consists in a secular drift and a 1yr-period term. This periodic term corresponds to the parallax due the Earth's revolution and the difference in distance between the ephemerides. 

By using the classical weighting scheme (with the weights given by \cite{Chesley2010}), we have a very similar orbit between NIMA 
and OrbFit (corresponding to AstDys ephemeris). 

Compared to other ephemerides, the advantages of the NIMA ephemeris are to allow for the use of more observations, not only MPC 
observations and to have the control of the weighting process.

\subsection{Offset variation}
\begin{figure}[h!] 
\centering 
\includegraphics[width=\columnwidth]{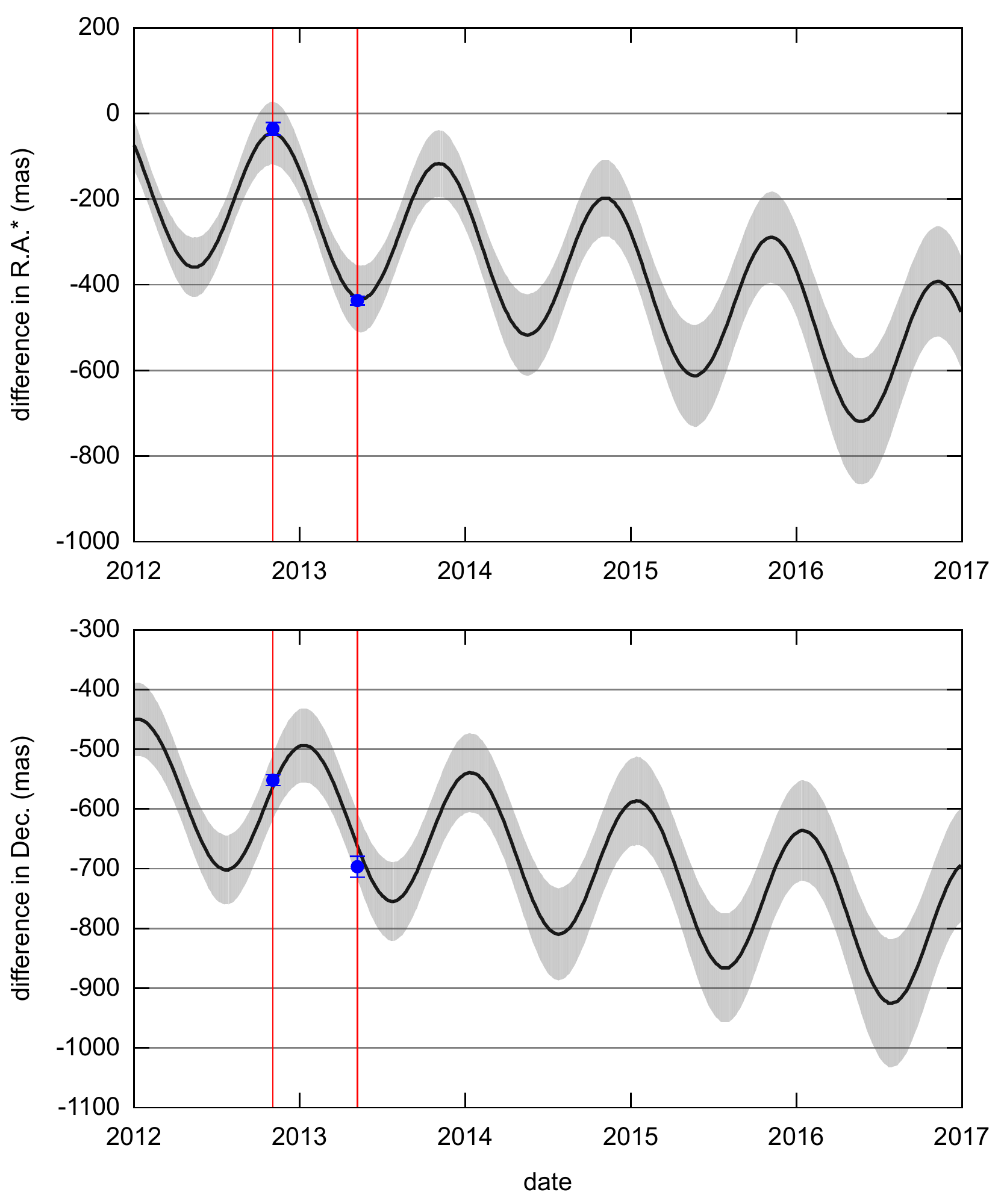}
\caption{Difference between NIMA and JPL ephemerides for 2004NT33 in right ascension (weighted by $\cos \delta$) and declination during 2012-2017. The grey area represents the uncertainty of NIMA ephemeris and the blue bullets and their error bars represent the offset observations used for NIMA ephemeris.}\label{F:2004NT33}
\end{figure}

By comparing the difference between NIMA ephemeris, that has been fitted to all positions (MPC + Offset observations) and JPL 
ephemeris detemined with only MPC positions, we have an estimation of the offset variation. Figure \ref{F:2004NT33} represents 
the difference between NIMA and JPL ephemerides for 2004NT33. For this object, we have two additional sets of offset observations made at 
ESO in November 2012 and May 2013. {The sets of observations correspond to one single night of observations where several observations were performed. The blue dots represent the average positions of each set and the error bar represents the standard deviation (1-$\sigma$)}. Obviously, NIMA ephemeris fits to 
ESO observations whereas JPL ephemeris does not. As commented in the previous section, the offset variation consists in a secular 
drift and a 1-year periodic term corresponding to a difference in the heliocentric distance between the two orbits. As expected, 
the offset is not constant with time. We can notice that the extrema of the oscillations correspond to the quadrature of the object 
(when elongation is 90 degrees).  The grey area in the figure represents the estimated uncertainty (1-$\sigma$) of the NIMA ephemeris 
and we can notice that the uncertainty increases with the time.

\subsection{Predictions of occultations}
\begin{figure}[h!] 
\centering 
\includegraphics[width=\columnwidth]{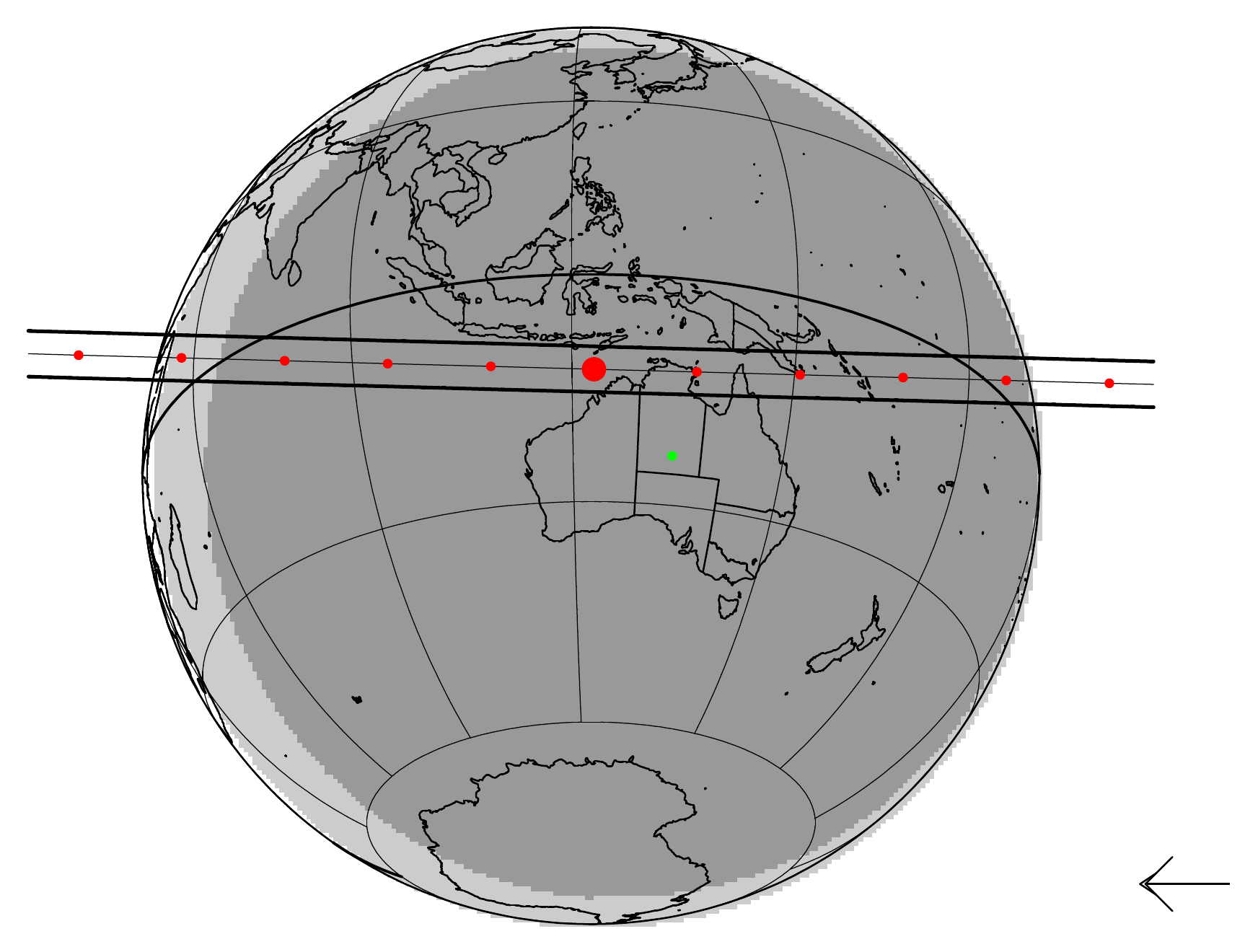}
\includegraphics[width=\columnwidth]{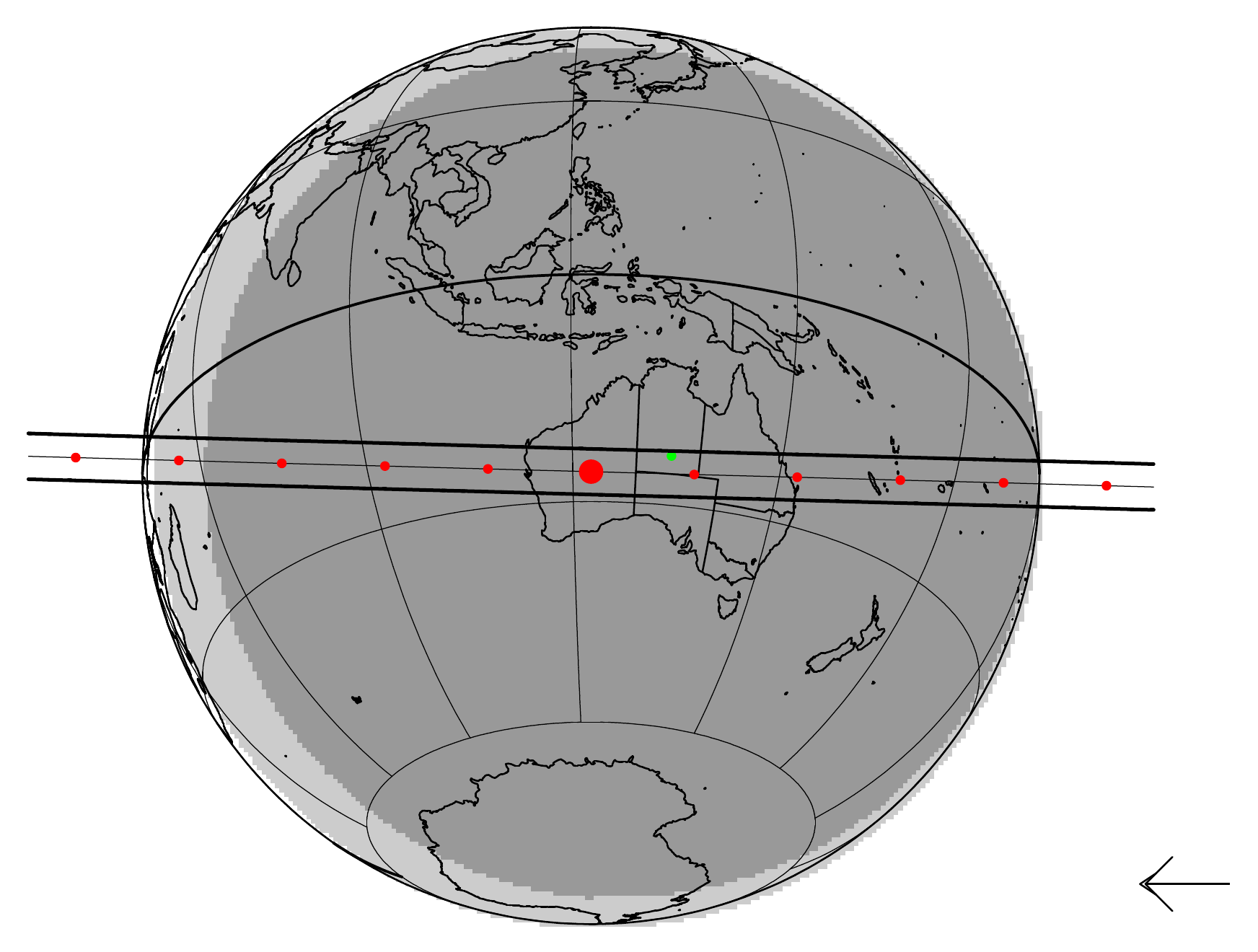}
\caption{Prediction map of the occultation by Ixion on 24 June 2014 with the offset method (top) and with the NIMA ephemeris (bottom). The occultation was detected in the place indicated by the green point.}\label{F:mapIxion}
\end{figure}

By design, the NIMA ephemeris is most suited for occultation predictions, as its better accuracy allows for more confident predictions farther in advance, for occultations to occur several months after the offset observations. We compare two predictions for the occultation by (28978) Ixion on 24 June 2014 with the offset method and with the NIMA ephemeris. Figure~\ref{F:mapIxion} shows the different predictions. The path of the shadow crosses over the North of Australia for the offset method and the Centre of Australia for the NIMA ephemeris. Actually, the occultation was successfully observed in the Centre of Australia (indicated by the green point), indicating that the prediction with NIMA ephemeris was more accurate. In fact, since July 2013 when NIMA ephemeris was applied to the occultation predictions until February 2015, 10 TNOs and 3 Centaurs events have been detected whereas 13 occultations by TNOs have been detected between 2009 and 2012. 

\subsection{Precise positions from occultations}\label{Ss:resultsocc}
As explained in Sect.~\ref{Ss:occ}, precise astrometric positions of TNOs can be deduced from previous positive occultations. We have reduced 14 astrometric positions for 8 different objects so far. Table~\ref{T:occ} presents the residuals in right ascension weighted by $\cos \delta$ (RA*) and in declination of these positions. Most of the positions have a precision less than 50~mas.    

\begin{table}[h!]
\begin{center}
\caption{Residuals of 14 astrometric positions deduced from previous occultations. The table indicates the name of the object, the date of the occultation, the weight used in orbit determination in mas, and the residuals in arcsec in RA* and Dec. The weight depends mostly on the quality of star position determination.}\label{T:occ}
\begin{tabular}{lcrrr}
\hline
\hline
Name & Date & Weight & \multicolumn{2}{c}{Residuals} \\
  &   &    & RA* & Dec \\
\hline
\hline
 2002KX14 & 2012-04-26 & 90 &  -0.057 & -0.003 \\
\hline
 2003AZ84 & 2011-01-08 & 40 &  -0.017 & -0.009 \\
          & 2012-02-03 & 40 &  -0.006 &  0.011 \\
\hline
 2003VS2  & 2014-03-04 & 75 &   0.004 & -0.015 \\
\hline
 Chariklo & 2013-06-03 & 40 &  -0.006 & -0.020 \\
          & 2014-02-16 & 40 &   0.006 &  0.039 \\
          & 2014-04-29 & 40 &   0.029 & -0.011 \\
          & 2014-06-28 & 40 &  -0.030 &  0.009 \\
\hline
 Eris 	  & 2013-08-29 & 75 &   0.006 &  0.007 \\
\hline
 Makemake & 2011-04-23 & 75 &  -0.010 &  0.083 \\
\hline
 Quaoar   & 2011-05-04 & 40 &   0.005 & -0.008 \\
          & 2012-02-17 & 40 &   0.000 & -0.033 \\
          & 2012-10-15 & 40 &  -0.003 & -0.002 \\
\hline
 Varuna   & 2013-01-08 & 40 &   0.014 & -0.005 \\
\hline
\hline
\end{tabular}
\end{center}
\end{table}

\begin{figure*}[h!] 
\centering 
\includegraphics[width=\columnwidth]{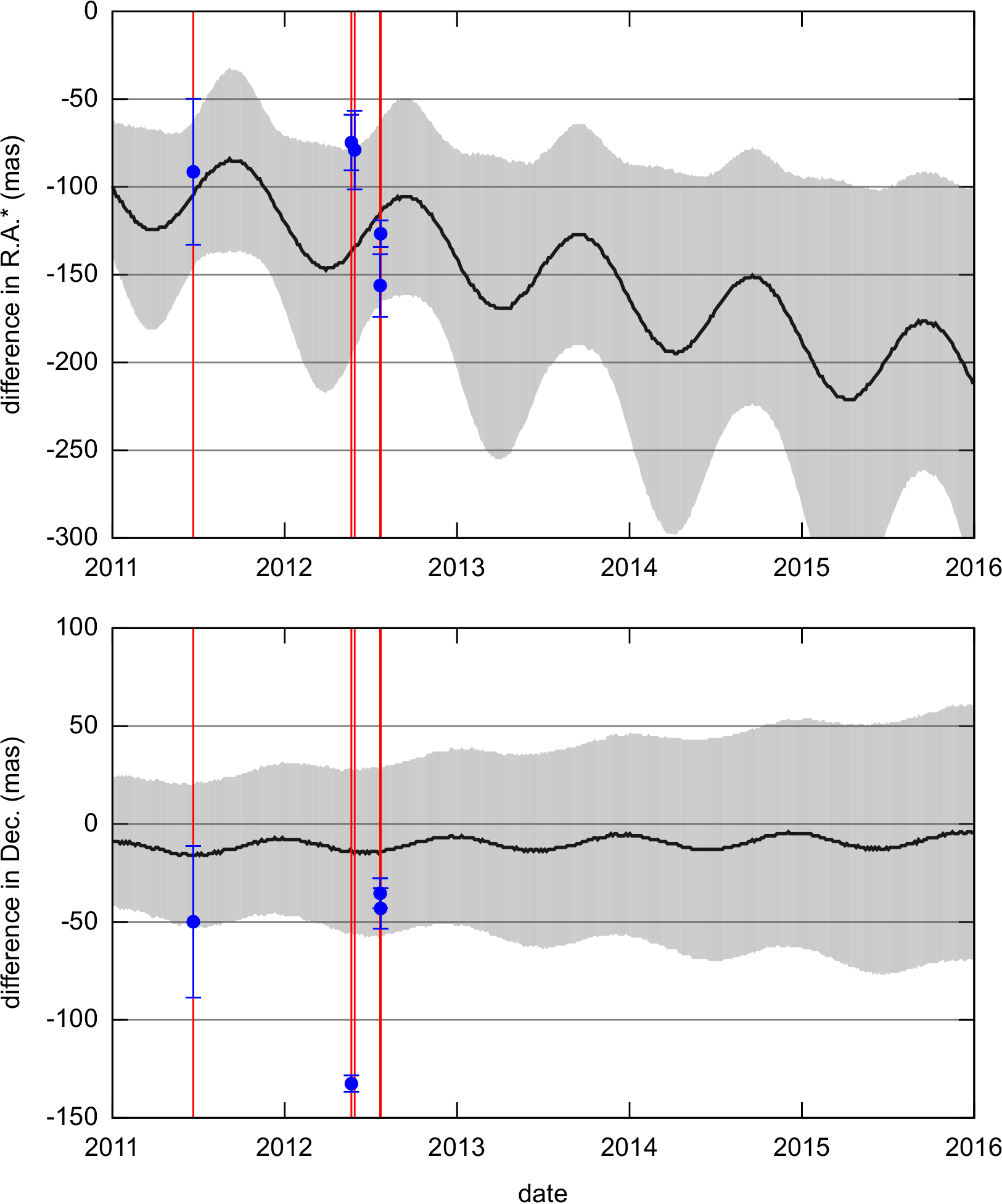}
\includegraphics[width=\columnwidth]{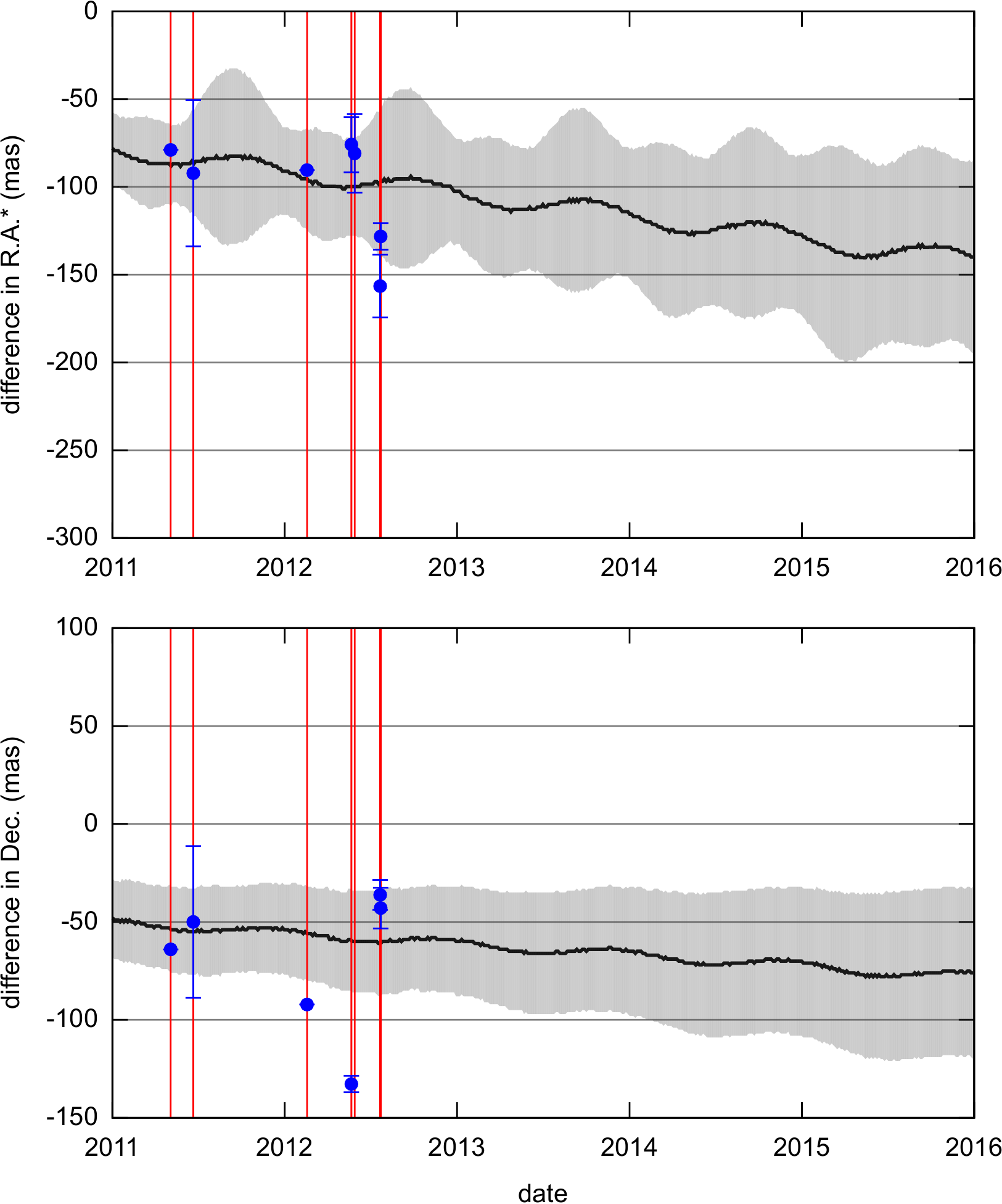}
\caption{Difference between NIMA and JPL ephemerides for (50000) Quaoar in right ascension weighted by $\cos \delta$ (top) and declination (bottom) during 2011-2016, by using only the positions until October 2012 (left) and by using the positions until October 2012 and plus the two positions from the two previous occultations in May 2011 and February 2012 (right). The grey aera represents the uncertainty of NIMA ephemeris and the blue bullets and their error bars represent the positions from the offset observations used for NIMA ephemeris.}\label{F:Quaoar}
\end{figure*}

\begin{figure*}[h!] 
\centering 
\includegraphics[width=\columnwidth]{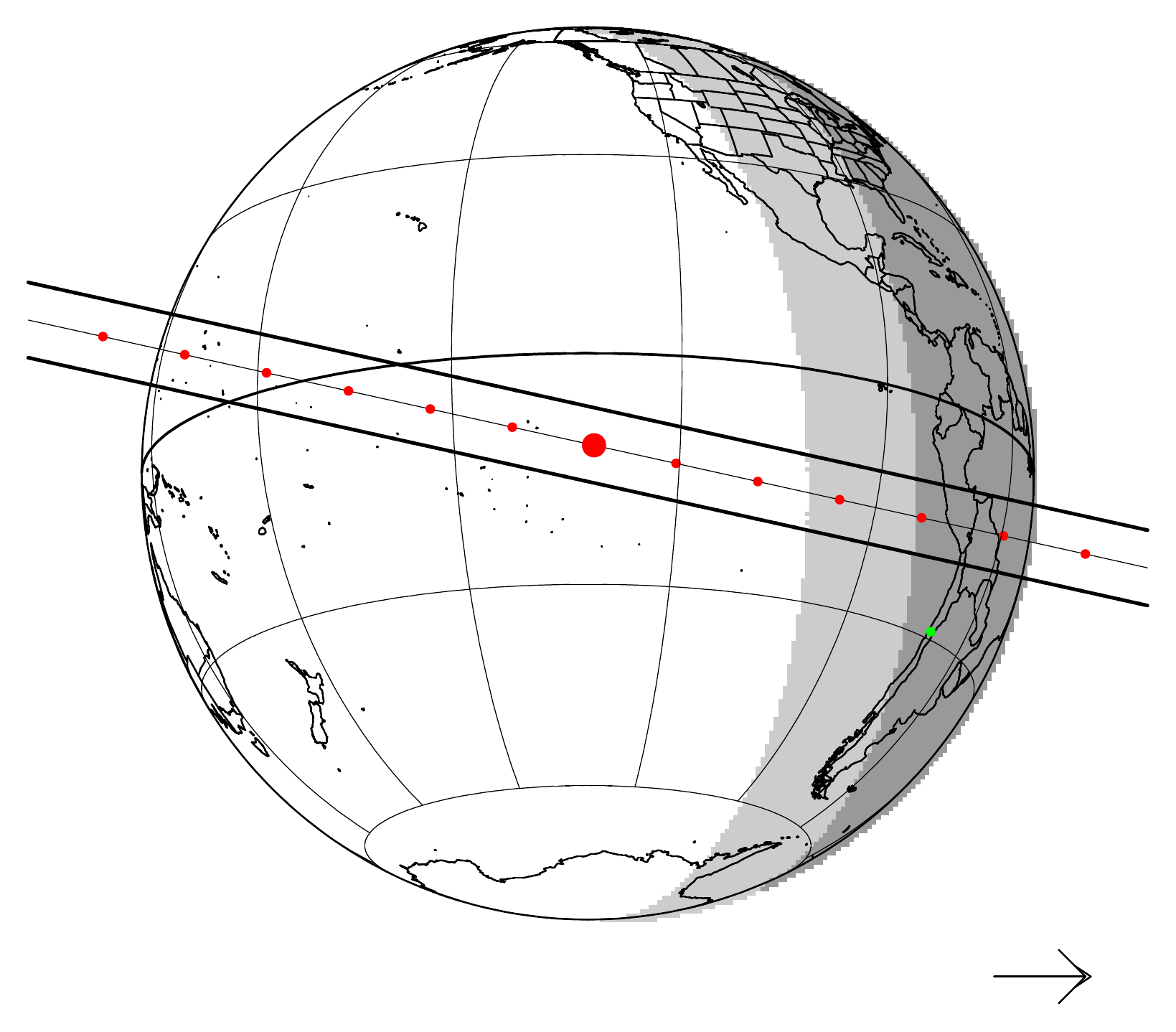}
\includegraphics[width=\columnwidth]{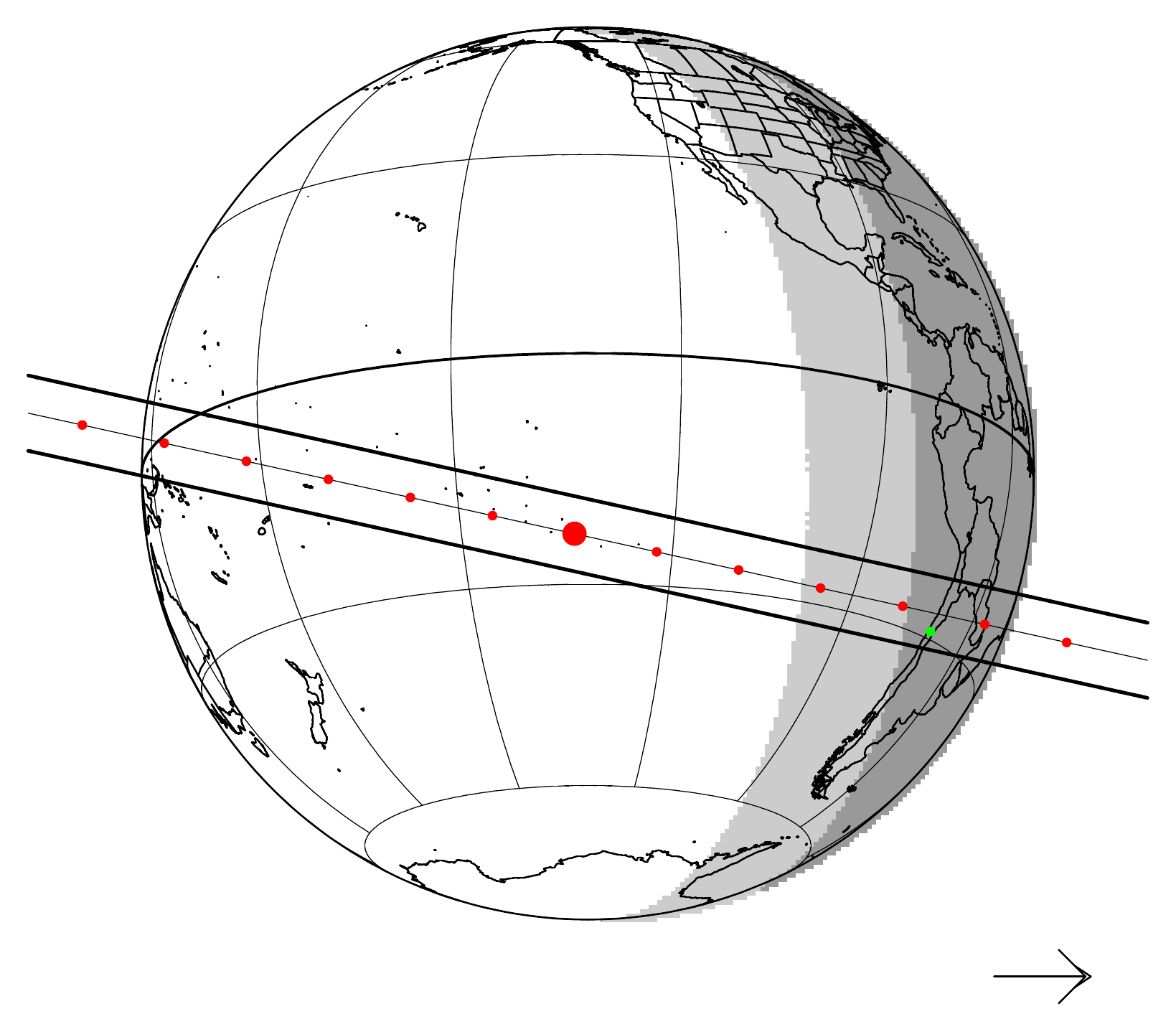}
\caption{Prediction map of the occultation by Quaoar on 15 October 2012 using only observations (left) and using observations and 2 previous occultations (right).}\label{F:mapQuaoar}
\end{figure*}

These positions can be used to refine the orbit and to improve the prediction of stellar occultations. For example, three positive occultations by (50000) Quaoar have been observed in 2011 and 2012 \citep{2013ApJ...773...26B}. To show that the positions provided by these occultations help to improve the orbit quality, we present the prediction of the third observed event, on 15 October 2012, with two different sets of positions. The first, using all the available observations until this date (including offset observations); and the second using the same observations plus the two previous positions deduced from the previous detected occultations, on May 2011 and February 2012. We created the prediction maps for the two cases, see Figure~\ref{F:mapQuaoar}. Figure~\ref{F:Quaoar} shows the difference between NIMA and JPL ephemerides in right ascension (weighted by $\cos \delta$) and declination for these two cases. The difference between the two orbits is about 50 mas in declination at the date of the occultation. For the configuration of this event, an uncertainty in right ascension corresponds to mainly to an uncertainty in time of occultation, whereas an uncertainty in declination corresponds to an uncertainty in the location of the path, which is more important for observation purposes. The two orbits correspond to two different paths of the shadow for the occultation on 15 October 2012 (Fig.~\ref{F:mapQuaoar}). The first one using only the observations predicts an occultation by Quaoar in the South of Peru whereas the second orbit using observations and positions of occultations predicts an occultation over Chile. Finally, the occultation was positively detected at Cerro Tololo by the PROMPT telescopes, in the Centre of Chile (green point on the figure), showing that the prediction was better thanks to the position given by previous occultations.

\subsection{Discussion}

As the position deduced from occultations is only affected by the error on the position of the occulted star, the derived position is more accurate than classical observations. In particular, these positions can highlight systematic errors in observations. Figure~\ref{F:Quaoar} reveals systematic errors in the positions from the two last sets of observations performed on May and July 2012. Even if the positions of previous occultations are taken into account, the orbit cannot match accurately these observed positions. The reason comes from the quality of all the positions used to make the orbit. The change of 80 mas in right ascension and 90 mas in declination between May and July 2012 may only be explained by systematic errors. They may come from zonal errors in stellar catalogue, the telescope, the sky conditions, etc.  

As a comparison, the positions from the offset observations on Fig.~\ref{F:2004NT33} may also be affected by systematic errors but since there is a few number of observations for 2004NT33, the two additional positions fit well the NIMA orbit (black line). In that case, systematic errors are hard to detect and we can only trust the observations. The main difference of this study is that the orbit determination as well as the uncertainty of the ephemeris now take into account possible systematic errors in the positions through the adopted weighting scheme.  

Systematic errors in the positions are currently the main limiting factor for accurate orbit determination. 
Some systematic errors are linked to observation such as the zonal error in the catalogue or the differential chromatic refraction, and they may be partially corrected thanks to the Gaia catalogue. Other are linked to dynamics such as the difference between the positions of the photocentre and the barycentre for binary systems, and may be only corrected with a careful modelling of the mutual orbit of the two bodies.

The theory of orbit determination allows to deal with positions with different precisions by using a weighting scheme but not with systematic errors.
Until systematic errors in positions from CCD images can be brought to a minimum, thanks to the astrometry from Gaia, the weighting scheme used in this study allows to partially deal with these errors. 

\subsection{TNO's ephemerides and observations}
As output of this study, we make available the ephemeris of the 51 selected TNOs and Centaurs during the period 2010-2020. Used to make predictions of stellar occultations, these ephemerides are available in bsp file usable with the SPICE library \citep{Acton1996} at the address \url{http://www.imcce.fr/~desmars/research/tno/}. These ephemerides will be regularly updated once new observations become available.

The predictions of forthcoming stellar occultations are available at \url{http://devel2.linea.gov.br/~braga.ribas/campaigns/} or \url{http://www.lesia.obspm.fr/perso/bruno-sicardy/}. For comparisons, predictions for past occultations are available at \url{http://devel2.linea.gov.br/~braga.ribas/campaigns/old.html}.

The offset observations of the selected TNOs are available on the CDS.
The statistics of the residuals, the number and the time-span of  MPC observations and offset observations are given in Table~\ref{T:residuals}. The offset observations have a better quality than MPC positions and generally they help to extend the period of positions.

\section{Conclusion}\label{S:conclusion}
The prediction of stellar occultations by TNOs is, and will be for a long time, thoroughly dependent 
on observations. Although the astrometry from the GAIA space mission will provide star positions with an accuracy from few microarcseconds to hundreds of microarcseconds, observations aiming at the determination of positions of TNOs will still be necessary for precise orbit calculation purposes. In this context, a large contribution is expected from deep-sky surveys, such as Pan-STARRS\footnote{Panoramic Survey Telescope and Rapid Response System : \url{http://pan-starrs.ifa.hawaii.edu}} or the LSST\footnote{Large Synoptic Survey Telescope : \url{http://www.lsst.org/lsst/}}. This 
survey will repeatedly observe the sky southern up to $\delta=+10$ degrees in 6 bands, also providing 
multiple observations for tens of thousands of TNOs. NIMA is a suitable tool to ingest these data and
provide improved ephemerides for these objects.

On the other hand, the astrometry and photometry from GAIA will greatly improve the astrometric reduction
of CCD and photographic images as well as renew the importance of old epoch images. In fact, old plates 
with solar system objects may be reduced with reference star positions having sub-mas accuracies at the 
plates' epoch. These old epoch positions are of utmost importance to the accuracy of ephemerides for
objects with long (hundreds of years) periods. In this context, it should be mentioned that our team 
has an image database with solar system objects, acquired at the Pico dos Dias Observatory, that 
spans about 20 years. Those with TNOs, from Pico dos Dias and La Silla, span half of this time. 
All of them will be re-reduced with the GAIA astrometric catalogue.

Another source of ephemeris improvement could also be obtained from the observational strategy.
Errors in the TNO's distance may amount to thousands of kilometres and reflect as one
year period oscillations in plots showing position differences between different ephemerides of
the same object. Here, it should be noted that observations are preferably made close to oppostion,
a configuration that is less sensitive to parallax effects in position when compared to quadrature.
Therefore, more frequent observations of TNOs in quadrature would improve the accuracy of their
ephemerides.

\begin{acknowledgements}
JD was supported by CNPq grant 161605/2012-5, JC acknowledges CNPq for a PQ2 fellowship (process number 308489/2013-6), FBR acknowledges PAPDRJ-FAPERJ/CAPES E-43/2013 number 144997, E-26/101.375/2014, RVM thanks grants CNPq-306885/2013, Capes/Cofecub-2506/2015, Faperj/PAPDRJ-45/2013. MA acknowledges CNPq grants 473002/2013-2, 482080/2009-4 and 308721/2011-0, and FAPERJ grant 111.488/2013. ARGJ thanks the financial support of CAPES. JLO acknowledges support from Proyecto de Excelencia de la Junta de Andalucía, J.A. 2012-FQM1776 and FEDER funds
\end{acknowledgements}


\bibliographystyle{aa} 
\bibliography{biblio} 

\begin{thebibliography}{20}
\expandafter\ifx\csname natexlab\endcsname\relax\def\natexlab#1{#1}\fi

\bibitem[{{Acton}(1996)}]{Acton1996}
{Acton}, C.~H. 1996, \planss, 44, 65

\bibitem[{{Assafin} {et~al.}(2012){Assafin}, {Camargo}, {Vieira Martins},
  {Braga-Ribas}, {Sicardy}, {Andrei}, \& {da Silva Neto}}]{Assafin2012}
{Assafin}, M., {Camargo}, J.~I.~B., {Vieira Martins}, R., {et~al.} 2012, \aap,
  541, A142

\bibitem[{{Braga-Ribas} {et~al.}(2013){Braga-Ribas}, {Sicardy}, {Ortiz},
  {Lellouch}, {Tancredi}, {Lecacheux}, {Vieira-Martins}, {Camargo}, {Assafin},
  {Behrend}, {Vachier}, {Colas}, {Morales}, {Maury}, {Emilio}, {Amorim},
  {Unda-Sanzana}, {Roland}, {Bruzzone}, {Almeida}, {Rodrigues}, {Jacques},
  {Gil-Hutton}, {Vanzi}, {Milone}, {Schoenell}, {Salvo}, {Almenares}, {Jehin},
  {Manfroid}, {Sposetti}, {Tanga}, {Klotz}, {Frappa}, {Cacella}, {Colque},
  {Neves}, {Alvarez}, {Gillon}, {Pimentel}, {Giacchini}, {Roques}, {Widemann},
  {Magalh{\~a}es}, {Thirouin}, {Duffard}, {Leiva}, {Toledo}, {Capeche},
  {Beisker}, {Pollock}, {Cede{\~n}o Monta{\~n}a}, {Ivarsen}, {Reichart},
  {Haislip}, \& {Lacluyze}}]{2013ApJ...773...26B}
{Braga-Ribas}, F., {Sicardy}, B., {Ortiz}, J.~L., {et~al.} 2013, \apj, 773, 26

\bibitem[{{Braga-Ribas} {et~al.}(2014{\natexlab{a}}){Braga-Ribas}, {Sicardy},
  {Ortiz}, {Snodgrass}, {Roques}, {Vieira-Martins}, {Camargo}, {Assafin},
  {Duffard}, {Jehin}, {Pollock}, {Leiva}, {Emilio}, {Machado}, {Colazo},
  {Lellouch}, {Skottfelt}, {Gillon}, {Ligier}, {Maquet}, {Benedetti-Rossi},
  {Gomes}, {Kervella}, {Monteiro}, {Sfair}, {El Moutamid}, {Tancredi},
  {Spagnotto}, {Maury}, {Morales}, {Gil-Hutton}, {Roland}, {Ceretta}, {Gu},
  {Wang}, {Harps{\o}e}, {Rabus}, {Manfroid}, {Opitom}, {Vanzi}, {Mehret},
  {Lorenzini}, {Schneiter}, {Melia}, {Lecacheux}, {Colas}, {Vachier},
  {Widemann}, {Almenares}, {Sandness}, {Char}, {Perez}, {Lemos}, {Martinez},
  {J{\o}rgensen}, {Dominik}, {Roig}, {Reichart}, {Lacluyze}, {Haislip},
  {Ivarsen}, {Moore}, {Frank}, \& {Lambas}}]{2014Natur.508...72B}
{Braga-Ribas}, F., {Sicardy}, B., {Ortiz}, J.~L., {et~al.} 2014{\natexlab{a}},
  \nat, 508, 72

\bibitem[{{Braga-Ribas} {et~al.}(2014{\natexlab{b}}){Braga-Ribas},
  {Vieira-Martins}, {Assafin}, {Camargo}, {Sicardy}, \&
  {Ortiz}}]{braga-ribas14b}
{Braga-Ribas}, F., {Vieira-Martins}, R., {Assafin}, M., {et~al.}
  2014{\natexlab{b}}, in Revista Mexicana de Astronomia y Astrofisica, vol. 27,
  Vol.~44, Revista Mexicana de Astronomia y Astrofisica Conference Series, 3--3

\bibitem[{{Camargo} {et~al.}(2014){Camargo}, {Vieira-Martins}, {Assafin},
  {Braga-Ribas}, {Sicardy}, {Desmars}, {Andrei}, {Benedetti-Rossi}, \&
  {Dias-Oliveira}}]{Camargo2014}
{Camargo}, J.~I.~B., {Vieira-Martins}, R., {Assafin}, M., {et~al.} 2014, \aap,
  561, A37

\bibitem[{{Carpino} {et~al.}(2003){Carpino}, {Milani}, \&
  {Chesley}}]{Carpino2003}
{Carpino}, M., {Milani}, A., \& {Chesley}, S.~R. 2003, \icarus, 166, 248

\bibitem[{{Chesley} {et~al.}(2010){Chesley}, {Baer}, \& {Monet}}]{Chesley2010}
{Chesley}, S.~R., {Baer}, J., \& {Monet}, D.~G. 2010, \icarus, 210, 158

\bibitem[{{Desmars}(2015)}]{Desmars2015}
{Desmars}, J. 2015, \aap, 575, A53

\bibitem[{{Desmars} {et~al.}(2009){Desmars}, {Arlot}, {Arlot}, {Lainey}, \&
  {Vienne}}]{Desmars2009b}
{Desmars}, J., {Arlot}, S., {Arlot}, J.-E., {Lainey}, V., \& {Vienne}, A. 2009,
  \aap, 499, 321

\bibitem[{{Elliot} {et~al.}(2010){Elliot}, {Person}, {Zuluaga}, {Bosh},
  {Adams}, {Brothers}, {Gulbis}, {Levine}, {Lockhart}, {Zangari}, {Babcock},
  {Dupr{\'e}}, {Pasachoff}, {Souza}, {Rosing}, {Secrest}, {Bright}, {Dunham},
  {Sheppard}, {Kakkala}, {Tilleman}, {Berger}, {Briggs}, {Jacobson}, {Valleli},
  {Volz}, {Rapoport}, {Hart}, {Brucker}, {Michel}, {Mattingly},
  {Zambrano-Marin}, {Meyer}, {Wolf}, {Ryan}, {Ryan}, {Morzinski}, {Grigsby},
  {Brimacombe}, {Ragozzine}, {Montano}, \& {Gilmore}}]{2010Natur.465..897E}
{Elliot}, J.~L., {Person}, M.~J., {Zuluaga}, C.~A., {et~al.} 2010, \nat, 465,
  897

\bibitem[{{Everhart}(1985)}]{Everhart1985}
{Everhart}, E. 1985, in Dynamics of Comets: Their Origin and Evolution,
  Proceedings of IAU Colloq. 83, held in Rome, Italy, June 11-15, 1984. Edited
  by Andrea Carusi and Giovanni B. Valsecchi. Dordrecht: Reidel, Astrophysics
  and Space Science Library. Volume 115, 1985, p.185, ed. {A.~Carusi \&
  G.~B.~Valsecchi}, 185

\bibitem[{{Folkner} {et~al.}(2014){Folkner}, {Williams}, {Boggs}, {Park}, \&
  {Kuchynka}}]{DE431}
{Folkner}, W., {Williams}, J., {Boggs}, D., {Park}, R., \& {Kuchynka}, P. 2014,
  JPL IPN Progress Reports, 42-196,
  \url{http://ipnpr.jpl.nasa.gov/progress_report/42-196/196C.pdf}

\bibitem[{{Fraser} {et~al.}(2013){Fraser}, {Gwyn}, {Trujillo}, {Stephens},
  {Kavelaars}, {Brown}, {Bianco}, {Boyle}, {Brucker}, {Hetherington}, {Joner},
  {Keel}, {Langill}, {Lister}, {McMillan}, \& {Young}}]{Fraser2013}
{Fraser}, W.~C., {Gwyn}, S., {Trujillo}, C., {et~al.} 2013, \pasp, 125, 1000

\bibitem[{{Lainey} {et~al.}(2004){Lainey}, {Duriez}, \& {Vienne}}]{Lainey2004a}
{Lainey}, V., {Duriez}, L., \& {Vienne}, A. 2004, \aap, 420, 1171

\bibitem[{{Monet} {et~al.}(2003){Monet}, {Levine}, {Canzian}, {Ables}, {Bird},
  {Dahn}, {Guetter}, {Harris}, {Henden}, {Leggett}, {Levison}, {Luginbuhl},
  {Martini}, {Monet}, {Munn}, {Pier}, {Rhodes}, {Riepe}, {Sell}, {Stone},
  {Vrba}, {Walker}, {Westerhout}, {Brucato}, {Reid}, {Schoening}, {Hartley},
  {Read}, \& {Tritton}}]{USNOB}
{Monet}, D.~G., {Levine}, S.~E., {Canzian}, B., {et~al.} 2003, \aj, 125, 984

\bibitem[{{Ortiz} {et~al.}(2012){Ortiz}, {Sicardy}, {Braga-Ribas},
  {Alvarez-Candal}, {Lellouch}, {Duffard}, {Pinilla-Alonso}, {Ivanov},
  {Littlefair}, {Camargo}, {Assafin}, {Unda-Sanzana}, {Jehin}, {Morales},
  {Tancredi}, {Gil-Hutton}, {de La Cueva}, {Colque}, {da Silva Neto},
  {Manfroid}, {Thirouin}, {Guti{\'e}rrez}, {Lecacheux}, {Gillon}, {Maury},
  {Colas}, {Licandro}, {Mueller}, {Jacques}, {Weaver}, {Milone}, {Salvo},
  {Bruzzone}, {Organero}, {Behrend}, {Roland}, {Vieira-Martins}, {Widemann},
  {Roques}, {Santos-Sanz}, {Hestroffer}, {Dhillon}, {Marsh}, {Harlingten},
  {Campo Bagatin}, {Alonso}, {Ortiz}, {Colazo}, {Lima}, {Oliveira}, {Kerber},
  {Smiljanic}, {Pimentel}, {Giacchini}, {Cacella}, \&
  {Emilio}}]{2012Natur.491..566O}
{Ortiz}, J.~L., {Sicardy}, B., {Braga-Ribas}, F., {et~al.} 2012, \nat, 491, 566

\bibitem[{{Sicardy} {et~al.}(2011){Sicardy}, {Ortiz}, {Assafin}, {Jehin},
  {Maury}, {Lellouch}, {Hutton}, {Braga-Ribas}, {Colas}, {Hestroffer},
  {Lecacheux}, {Roques}, {Santos-Sanz}, {Widemann}, {Morales}, {Duffard},
  {Thirouin}, {Castro-Tirado}, {Jel{\'{\i}}nek}, {Kub{\'a}nek}, {Sota},
  {S{\'a}nchez-Ram{\'{\i}}rez}, {Andrei}, {Camargo}, {da Silva Neto}, {Gomes},
  {Martins}, {Gillon}, {Manfroid}, {Tozzi}, {Harlingten}, {Saravia}, {Behrend},
  {Mottola}, {Melendo}, {Peris}, {Fabregat}, {Madiedo}, {Cuesta}, {Eibe},
  {Ull{\'a}n}, {Organero}, {Pastor}, {de Los Reyes}, {Pedraz}, {Castro}, {de La
  Cueva}, {Muler}, {Steele}, {Cebri{\'a}n},
  {Monta{\~n}{\'e}s-Rodr{\'{\i}}guez}, {Oscoz}, {Weaver}, {Jacques}, {Corradi},
  {Santos}, {Reis}, {Milone}, {Emilio}, {Guti{\'e}rrez}, {V{\'a}zquez}, \&
  {Hern{\'a}ndez-Toledo}}]{2011Natur.478..493S}
{Sicardy}, B., {Ortiz}, J.~L., {Assafin}, M., {et~al.} 2011, \nat, 478, 493

\bibitem[{{Widemann} {et~al.}(2009){Widemann}, {Sicardy}, {Dusser}, {Martinez},
  {Beisker}, {Bredner}, {Dunham}, {Maley}, {Lellouch}, {Arlot}, {Berthier},
  {Colas}, {Hubbard}, {Hill}, {Lecacheux}, {Lecampion}, {Pau}, {Rapaport},
  {Roques}, {Thuillot}, {Hills}, {Elliott}, {Miles}, {Platt}, {Cremaschini},
  {Dubreuil}, {Cavadore}, {Demeautis}, {Henriquet}, {Labrevoir}, {Rau},
  {Coliac}, {Piraux}, {Marlot}, {Marlot}, {Gorry}, {Sire}, {Bayle}, {Simian},
  {Blommers}, {Fulgence}, {Leyrat}, {Sauzeaud}, {Stephanus}, {Rafaelli},
  {Buil}, {Delmas}, {Desnoux}, {Jasinski}, {Klotz}, {Marchais}, {Rieugni{\'e}},
  {Bouderand}, {Cazard}, {Lambin}, {Pujat}, {Schwartz}, {Burlot}, {Langlais},
  {Rivaud}, {Brochard}, {Dupouy}, {Lavayssi{\`e}re}, {Chaptal}, {Daiffallah},
  {Clarasso-Llauger}, {Aloy Dom{\'e}nech}, {Gabald{\'a}-S{\'a}nchez},
  {Otazu-Porter}, {Fern{\'a}ndez}, {Masana}, {Ardanuy}, {Casas}, {Ros},
  {Casarramona}, {Schnabel}, {Roca}, {Labordena}, {Canales-Moreno}, {Ferrer},
  {Rivas}, {Ortiz}, {Fern{\'a}ndez-Arozena}, {Mart{\'{\i}}n-Rodr{\'{\i}}guez},
  {Cidad{\~a}o}, {Coelho}, {Figuereido}, {Gon{\c c}alves}, {Marciano}, {Nunes},
  {R{\'e}}, {Saraiva}, {Tonel}, {Cl{\'e}rigo}, {Oliveira}, {Reis},
  {Ewen-Smith}, {Ward}, {Ford}, {Gon{\c c}alves}, {Porto}, {Laurindo Sobrinho},
  {Teodoro de Gois}, {Joaquim}, {Afonso da Silva Mendes}, {van Ballegoij},
  {Jones}, {Callender}, {Sutherland}, {Bumgarner}, {Imbert}, {Mitchell},
  {Lockhart}, {Barrow}, {Cornwall}, {Arnal}, {Eleizalde}, {Valencia}, {Ladino},
  {Lizardo}, {Guill{\'e}n}, {S{\'a}nchez}, {Pe{\~n}a}, {Radaelli}, {Santiago},
  {Vieira}, {Mendt}, {Rosenzweig}, {Naranjo}, {Contreras}, {D{\'{\i}}az},
  {Guzm{\'a}n}, {Moreno}, {Omar Porras}, {Recalde}, {Mascar{\'o}}, {Birnbaum},
  {C{\'o}sias}, {L{\'o}pez}, {Pallo}, {Percz}, {Pulupa}, {Simba{\~n}a},
  {Yajam{\'{\i}}n}, {Rodas}, {Denzau}, {Kretlow}, {Vald{\'e}s Sada},
  {Hern{\'a}ndez}, {Hern{\'a}ndez}, {Wilson}, {Castro}, \&
  {Winkel}}]{2009Icar..199..458W}
{Widemann}, T., {Sicardy}, B., {Dusser}, R., {et~al.} 2009, \icarus, 199, 458

\bibitem[{{Zacharias} {et~al.}(2013){Zacharias}, {Finch}, {Girard}, {Henden},
  {Bartlett}, {Monet}, \& {Zacharias}}]{UCAC4}
{Zacharias}, N., {Finch}, C.~T., {Girard}, T.~M., {et~al.} 2013, \aj, 145, 44

\end{thebibliography}

\onecolumn
\begin{longtable}{lrrrrrc}
\caption{Statistics of post-fit residuals for the 51 selected TNO and Centaur for MPC positions (first line) and for the positions from the offset observations (second line) used for orbit determination. The mean $\mu$ and the standard deviation $\sigma$ of right ascension weighted by $\cos \delta$ and of the declination are indicated as well as the number of accepted positions and the time span.}\label{T:residuals}  
\\ 
TNO & $\mu_{\alpha*}$ &  $\sigma_{\alpha*}$ & $\mu_{\delta}$ &  $\sigma_{\delta}$ & number & time-span \\
\hline
\hline
\endfirsthead
\caption{continued.}\\
TNO & $\mu_{\alpha*}$ &  $\sigma_{\alpha*}$ & $\mu_{\delta}$ &  $\sigma_{\delta}$ & number & time-span \\
\hline	
\hline	
\endhead
(24835) 1995SM55   &     -0.047 &      0.683 &      0.017 &      0.459 &     125 & 1982-2012  \\
                   &      0.006 &      0.019 &     -0.001 &      0.043 &      10 &    2012    \\
 \hline
(26375) 1999DE9    &      0.006 &      0.468 &     -0.052 &      0.294 &      71 & 1990-2008  \\
                   &     -0.022 &      0.056 &      0.009 &      0.033 &      40 & 2012-2013  \\
 \hline
(47171) 1999TC36   &     -0.034 &      0.519 &     -0.046 &      0.545 &     106 & 1974-2013  \\
                   &      0.010 &      0.030 &      0.014 &      0.019 &      37 & 2012-2013  \\
 \hline
(55565) 2002AW197  &      0.081 &      0.214 &      0.034 &      0.190 &     115 & 1997-2013  \\
                   &     -0.006 &      0.047 &      0.013 &      0.054 &      55 & 2012-2013  \\
 \hline
(119951) 2002KX14  &      0.081 &      0.196 &      0.030 &      0.218 &      56 & 1984-2011  \\
                   &      0.039 &      0.035 &     -0.017 &      0.030 &      36 & 2012-2013  \\
 \hline
(307261) 2002MS4   &      0.046 &      0.297 &      0.028 &      0.382 &      58 & 1954-2009  \\
                   &      0.002 &      0.046 &      0.011 &      0.031 &      47 & 2012-2014  \\
 \hline
(84522) 2002TC302  &      0.013 &      0.491 &      0.012 &      0.327 &      96 & 2000-2013  \\
                   &     -0.014 &      0.071 &      0.004 &      0.070 &     125 & 2011-2014  \\
 \hline
(55636) 2002TX300  &     -0.003 &      0.234 &      0.016 &      0.306 &     341 & 1954-2013  \\
                   &     -0.070 &      0.028 &      0.002 &      0.011 &      10 &    2013    \\
 \hline
(55637) 2002UX25   &     -0.058 &      0.342 &     -0.047 &      0.429 &      74 & 1991-2013  \\
                   &     -0.009 &      0.057 &      0.066 &      0.065 &      31 &    2012    \\
 \hline
(55638) 2002VE95   &     -0.008 &      0.254 &      0.002 &      0.278 &     193 & 1990-2013  \\
                   &     -0.018 &      0.044 &     -0.010 &      0.019 &      23 & 2012-2013  \\
 \hline
(119979) 2002WC19  &      0.036 &      0.257 &     -0.016 &      0.364 &      74 & 2001-2012  \\
                   &      0.018 &      0.061 &     -0.036 &      0.057 &      24 & 2012-2013  \\
 \hline
(208996) 2003AZ84  &      0.037 &      0.384 &      0.018 &      0.351 &     103 & 1996-2014  \\
                   &      0.012 &      0.062 &      0.009 &      0.067 &      79 & 2011-2013  \\
 \hline
(120132) 2003FY128 &      0.093 &      0.307 &      0.053 &      0.410 &      68 & 1989-2012  \\
                   &      0.001 &      0.035 &     -0.008 &      0.028 &      32 & 2012-2013  \\
 \hline
(120178) 2003OP32  &      0.004 &      0.437 &      0.034 &      0.356 &      68 & 1990-2011  \\
                   &     -0.003 &      0.028 &     -0.002 &      0.021 &      59 & 2012-2013  \\
 \hline
2003UZ41           &      0.111 &      0.322 &      0.123 &      0.418 &      36 & 1954-2010  \\
                   &     -0.008 &      0.017 &     -0.006 &      0.013 &      10 &    2012    \\
 \hline
(84922) 2003VS2    &     -0.024 &      0.294 &      0.038 &      0.364 &     177 & 1991-2014  \\
                   &      0.013 &      0.077 &     -0.020 &      0.050 &      37 & 2011-2014  \\
 \hline
(90568) 2004GV9    &      0.034 &      0.493 &      0.023 &      0.472 &      62 & 1954-2011  \\
                   &      0.001 &      0.012 &      0.006 &      0.013 &      18 &    2013    \\
 \hline
2004NT33           &      0.003 &      0.346 &      0.182 &      0.508 &      27 & 1982-2010  \\
                   &     -0.020 &      0.069 &      0.005 &      0.068 &      25 & 2011-2013  \\
 \hline
(175113) 2004PF115 &      0.067 &      0.326 &      0.099 &      0.323 &      37 & 1992-2010  \\
                   &     -0.005 &      0.012 &     -0.007 &      0.008 &      10 &    2012    \\
 \hline
(120348) 2004TY364 &     -0.029 &      0.312 &      0.091 &      0.404 &      20 & 1983-2005  \\
                   &     -0.001 &      0.015 &     -0.001 &      0.017 &      14 & 2012-2013  \\
 \hline
(144897) 2004UX10  &      0.059 &      0.352 &      0.149 &      0.572 &      83 & 1953-2007  \\
                   &      0.002 &      0.016 &     -0.028 &      0.020 &       8 &    2012    \\
 \hline
2011FX62-2005CC79  &      0.100 &      0.374 &      0.082 &      0.331 &      34 & 2002-2012  \\
                   &     -0.002 &      0.032 &     -0.012 &      0.025 &      29 & 2012-2013  \\
 \hline
(303775) 2005QU182 &     -0.082 &      0.217 &      0.008 &      0.304 &      81 & 1974-2011  \\
                   &      0.025 &      0.037 &     -0.027 &      0.053 &       9 &    2012    \\
 \hline
(145451) 2005RM43  &      0.004 &      0.200 &     -0.028 &      0.198 &     206 & 1976-2014  \\
                   &      0.008 &      0.022 &     -0.005 &      0.019 &      10 &    2012    \\
 \hline
(145452) 2005RN43  &     -0.014 &      0.171 &     -0.011 &      0.148 &     314 & 1954-2013  \\
                   &      0.019 &      0.012 &      0.015 &      0.008 &      10 &    2012    \\
 \hline
(145453) 2005RR43  &      0.009 &      0.193 &     -0.029 &      0.201 &     221 & 1976-2014  \\
                   &     -0.005 &      0.011 &      0.004 &      0.014 &      10 &    2012    \\
 \hline
(202421) 2005UQ513 &     -0.130 &      0.439 &     -0.021 &      0.355 &      63 & 1990-2013  \\
                   &      0.047 &      0.098 &      0.003 &      0.048 &      55 & 2012-2014  \\
 \hline
2007JH43           &      0.052 &      0.249 &      0.036 &      0.348 &      45 & 1984-2012  \\
                   &     -0.018 &      0.055 &     -0.009 &      0.026 &      47 & 2012-2013  \\
 \hline
(278361) 2007JJ43  &      0.041 &      0.280 &      0.159 &      0.194 &     104 & 2002-2012  \\
                   &     -0.005 &      0.015 &     -0.024 &      0.014 &      22 &    2013    \\
 \hline
(225088) 2007OR10  &      0.023 &      0.281 &      0.039 &      0.351 &      71 & 1985-2011  \\
                   &     -0.006 &      0.045 &      0.001 &      0.027 &      12 &    2012    \\
 \hline
(229762) 2007UK126 &     -0.016 &      0.215 &      0.053 &      0.272 &      73 & 1982-2013  \\
                   &     -0.001 &      0.012 &     -0.026 &      0.010 &      10 &    2012    \\
 \hline
2008OG19           &     -0.012 &      0.312 &     -0.050 &      0.495 &      27 & 2008-2012  \\
                   &      0.000 &      0.016 &      0.002 &      0.015 &      18 & 2012-2013  \\
 \hline
2010EK13           &      0.109 &      0.307 &      0.078 &      0.158 &     123 & 2002-2011  \\
                   &     -0.025 &      0.028 &     -0.042 &      0.055 &      30 & 2012-2013  \\
 \hline
(55576) Amycus     &      0.007 &      0.697 &     -0.027 &      0.514 &      76 & 1987-2007  \\
                   &     -0.023 &      0.058 &     -0.006 &      0.020 &      28 &    2013    \\
 \hline
(8405)  Asbolus    &     -0.010 &      0.533 &     -0.018 &      0.510 &     466 & 1995-2011  \\
                   &     -0.012 &      0.129 &     -0.011 &      0.109 &      13 & 2012-2013  \\
 \hline
(54598) Bienor     &     -0.024 &      0.413 &     -0.009 &      0.343 &     167 & 1975-2013  \\
                   &      0.011 &      0.037 &      0.017 &      0.052 &      69 & 2012-2014  \\
 \hline
(10199) Chariklo   &      0.058 &      0.562 &      0.027 &      0.476 &     571 & 1988-2011  \\
                   &     -0.010 &      0.074 &      0.009 &      0.039 &     336 & 2011-2014  \\
 \hline
(2060)  Chiron     &     -0.011 &      0.576 &     -0.033 &      0.593 &    1304 & 1895-2014  \\
                   &      0.088 &      0.152 &     -0.062 &      0.078 &      78 &    2014    \\
 \hline
(83982) Crantor    &     -0.027 &      0.380 &      0.011 &      0.423 &     116 & 2001-2014  \\
                   &     -0.001 &      0.008 &     -0.024 &      0.008 &       8 &    2013    \\
 \hline
(60558) Echeclus   &      0.053 &      0.553 &      0.060 &      0.490 &     686 & 1979-2014  \\
                   &         -- &         -- &         -- &         -- &       0 &     --     \\
 \hline
(136199) Eris      &      0.016 &      0.317 &      0.015 &      0.256 &     615 & 1954-2014  \\
                   &     -0.059 &      0.068 &     -0.004 &      0.074 &      96 & 2007-2012  \\
 \hline
(136108) Haumea    &      0.022 &      0.313 &      0.014 &      0.284 &    1139 & 1955-2014  \\
                   &      0.021 &      0.051 &     -0.001 &      0.104 &     148 & 2011-2013  \\
 \hline
(38628) Huya       &      0.013 &      0.500 &      0.044 &      0.497 &     151 & 1996-2014  \\
                   &     -0.008 &      0.018 &     -0.001 &      0.018 &      22 &    2013    \\
 \hline
(28978) Ixion      &     -0.022 &      0.319 &     -0.025 &      0.340 &     172 & 1982-2014  \\
                   &      0.026 &      0.045 &     -0.013 &      0.060 &     216 & 2009-2014  \\
 \hline
(136472) Makemake  &     -0.013 &      0.418 &      0.023 &      0.281 &    1081 & 1955-2014  \\
                   &      0.000 &      0.067 &     -0.058 &      0.070 &     484 & 2009-2013  \\
 \hline
(90482) Orcus      &      0.022 &      0.275 &     -0.012 &      0.201 &     434 & 1951-2014  \\
                   &     -0.034 &      0.045 &      0.006 &      0.047 &      58 & 2011-2014  \\
 \hline
(50000) Quaoar     &      0.004 &      0.387 &      0.025 &      0.344 &     400 & 1954-2014  \\
                   &     -0.003 &      0.043 &     -0.017 &      0.047 &     111 & 2011-2013  \\
 \hline
(120347) Salacia   &     -0.004 &      0.227 &      0.034 &      0.210 &      69 & 1982-2010  \\
                   &      0.008 &      0.013 &     -0.013 &      0.016 &      12 &    2012    \\
 \hline
(90377) Sedna      &     -0.084 &      0.536 &      0.101 &      0.523 &      91 & 1990-2012  \\
                   &      0.086 &      0.089 &     -0.044 &      0.044 &      23 &    2012    \\
 \hline
(174567) Varda     &     -0.006 &      0.332 &      0.049 &      0.354 &      72 & 1980-2010  \\
                   &     -0.006 &      0.032 &     -0.002 &      0.021 &      29 & 2013-2014  \\
 \hline
(20000) Varuna     &     -0.077 &      0.422 &      0.079 &      0.372 &     301 & 1954-2014  \\
                   &     -0.007 &      0.051 &     -0.009 &      0.028 &     172 & 2012-2013  \\
\hline
\hline
\end{longtable}

\end{document}